# 基于 LSTM-AE 的动态信道图谱构建*

高塬 [1]，谢文静 [1]，刘一鸣 [1]，郭馨雨 [1]，胡斌涛 [2]，杜剑波 [3]，徐树公 [2**]
（1.上海大学通信与信息工程学院，上海 200444；
2.西交利物浦大学智能工程学院，江苏 苏州 215123；
3.西安邮电大学通信与信息工程学院，陕西 西安 710121）
收稿日期：2025-11-30
*基金项目：上海市自然科学基金项目（25ZR1402148）“多点协作通信感知一体化网络关键技术研究”
**通信作者
【摘要】随着第六代（6G）移动通信系统的发展，CSI（Channel State Information，信道状态信息）是提升网络性至关重要的信息。传统的信道图谱（Channel Charting）方法通过将高维 CSI 数据映射到低维空间，从而揭示无线信道与物理环境之间的关系。然而，现有的信道图谱方法大多侧重于静态几何结构的学习，忽视了信道随时间变化的动态特性，导致在复杂动态环境中，信道图谱的稳定性和拓扑一致性较差。为了解决这一问题，本文提出了一种结合 LSTM（Long Short-Term Memory，长短时记忆网络）和 AE（Auto-Encoder，自编码器）的时序信道图谱构建方法（LSTM-AE-信道图谱），该方法在传统信道图谱框架的基础上融入了时序建模机制。通过引入 LSTM 网络捕捉 CSI 的时序依赖性，并使用 AE 学习低维的连续潜在表示，所提出的方法能够在保证信道几何一致性的同时，显式建模信道的时变特性。实验结果表明，所提出的方法在多个真实通信场景中均表现出了优异的性能，特别是在信道图谱的稳定性、轨迹连续性以及长期预测能力方面，相较于传统信道图谱方法，具有显著的优势。

# Dynamic Channel Charting: An LSTM-AE-based Approach

GAO Yuan[1], XIE Wenjing [1], LIU Yiming[1], GUO Xinyu[1], HU Bintao[2], DU Jianbo[3], XU Shugong[2]
(1. School of Communication and Information Engineering, Shanghai University, Shanghai 200444, China;
2. Department of Intelligent Science, Xi’an Jiaotong-Liverpool University, Suzhou 215123, China;
3. School of Communication and Information Engineering, Xi'an University of Posts and Telecommunications, Xi'an 710061, China)
[Abstract] With the development of the sixth-generation (6G) communication system, Channel State Information (CSI) plays a crucial role in improving network performance. Traditional Channel Charting (CC) methods map high-dimensional CSI data to low-dimensional spaces to help reveal the geometric structure of wireless channels. However, most existing CC methods focus on learning static geometric structures and ignore the dynamic nature of the channel over time, leading to instability and poor topological consistency of the channel charting in complex environments. To address this issue, this paper proposes a novel time-series channel charting approach based on the integration of Long Short-Term Memory (LSTM) networks and Auto encoders (AE) (LSTM-AE-CC). This method incorporates a temporal modeling mechanism into the traditional CC framework, capturing temporal dependencies in CSI using LSTM and learning continuous latent representations with AE. The proposed method ensures both geometric consistency of the channel and explicit modeling of the time-varying properties. Experimental results demonstrate that the proposed method outperforms traditional CC methods in various real-world communication scenarios, particularly in terms of channel charting stability, trajectory continuity, and long-term predictability.

## 0　引言

随着第六代（6G）移动通信系统的逐步演进，大规模 MIMO 及分布式天线技术被视为提升网络容量、覆盖范围和可靠性的关键支撑技术。其中，CSI（Channel State Information，信道状态信息）是实现链路自适应、预编码和资源调度的核心基础[1-2]。然而，在用户高速移动和复杂无线环境下，信道呈现出高度时变特性，导致 CSI 在获取后迅速失效，即所谓“信道老化（Channel Aging）”现象，从而严重制约系统性能[3]。为此，学术界开展了针对信道预测的广泛研究。传统方法多采用基于统计模型的信道预测技术，如 Kalman 滤波和 Wiener 预测等，这类方法依赖于对信道统计特性的精确建模，在复杂多径及非平稳场景中往往难以取得理想效果[1]。近年来，随着深度学习的发展，基于 RNN（Recurrent Neural Network，循环神经网络）、

LSTM（Long Short-Term Memory，长短时记忆网络）和 GRU（Gated Recurrent Unit，门控循环单元）的数据驱动方法被引入信道预测领域[4]，在一定程度上提高了预测精度。但这类方法通常直接针对 CSI 序列进行回归建模，忽略了无线信道中潜在的空间结构信息，难以刻画信道随用户运动产生的内在几何演化规律[5],[6]。

为了有效提取无线信道中的空间结构信息，信道图谱（Channel Charting）的概念被提出，它采用无监督学习的方式，将不同位置的 CSI 映射到低维潜空间，从而保留物理空间的邻域关系[5-6]。已有研究表明，信道图谱能够有效提升无线网络中无线定位[7]、波束管理[8]的性能。然而，现有信道图谱方法大多侧重于保持信道的"静态几何一致性"，即只关注样本间的空间邻近关系，而忽略信道随时间演化过程中的动态特性[9]。因此，在移动用户轨迹连续变化或环境快速切换时，信道图谱学到的嵌入结果往往缺乏时间一致性，导致信道图谱不稳定、拓扑漂移明显，从而限制了其在实际系统中的适用性。为提高信道图谱对时变信道的建模能力，部分研究尝试在信道图谱嵌入结果基础上引入插值或简单时序预测方法来进行信道外推[10]。然而，该类方法本质上仍是对已构建图表的后处理操作，依赖于流形的光滑假设，在复杂环境中容易产生误差累积，难以刻画用户运动所隐含的动力学规律。

基于上述问题，本文在信道图谱的构建引入无线信道在时间维度上的演化过程，构建具有动态一致性的信道潜空间。为此，本文提出一种基于 LSTM 与 AE（Auto-Encoder，自编码器）的时序信道图谱构建方法，将时间建模机制直接融入信道图谱的表示学习过程，使模型在学习低维嵌入的同时，显式刻画 CSI 的时序依赖关系，从而实现对动态信道结构的更高质量建模。

本文的主要贡献可总结为以下三个方面：

（1）提出一种面向动态信道的信道图谱建模范式：提出时空联合建模的新思路，将信道图谱从静态嵌入问题构造为动态表示学习问题，使信道图谱不仅具有空间一致性，同时具备时间稳定性。

（2）提出了一种基于 LSTM-AE 的信道图谱构建框架：通过 LSTM 对信道特征序列进行建模以刻画其时间演化过程，同时利用 AE 对高维信道特征进行非线性降维，学习低维嵌入表示以反映信道的几何结构特性，从而实现对信道空间结构与时间动态特性的联合建模。

（3）在真实轨迹场景下对所提出方法的有效性进行了测试：基于真实通信场景及用户运动轨迹数据开展实验，并与传统信道图谱插值方法及神经网络模型进行对比，实验结果表明，所提出的 LSTM-AE-信道图谱在嵌入稳定性、轨迹连续性以及长期预测性能等方面均表现出显著优势。

# 1 相关工作

## 1.1 基于 CSI 的时序预测研究现状

随着分布式大规模 MIMO 及 6G 系统的演进，CSI 在波束赋形、资源调度及通信链路自适应控制中发挥着核心作用。然而，在高速移动与复杂传播环境下，CSI 的时效性问题日益突出，限制了系统实时性和可靠性。因此，对 CSI 进行准确的预测，已成为当前无线通信领域的重要研究方向。

传统 CSI 预测方法主要依赖统计信道模型及经典时间序列分析工具，如 AR（Auto-Regressive，自回归模型）、卡尔曼滤波、LMMS（Linear Minimum Mean Square，线性最小均方预测器）等[1]。这类方法在平稳信道条件下具备一定有效性，但在多径变化剧烈或环境高度非平稳的场景中性能显著下降，难以满足新一代通信系统对精度与鲁棒性的要求。随着深度学习的发展，基于神经网络的 CSI 预测方法逐渐成为研究热点。研究者尝试利用 FNN（Feed-Forward Network，前馈神经网络）、CNN（Convolutional Neural Network，卷积神经网络）、RNN 等模型建立 CSI 的非线性时间演化映射。其中，LSTM 和 GRU 等门控循环网络因其在长期依赖建模方面的优势，被广泛应用于多步 CSI 的预测任务中[11-13]。相关研究表明，相较于传统预测器，深度模型在多步预测精度方面具有明显优势。

尽管上述方法在数值预测层面取得了一定进展，但其普遍存在以下不足：一是直接对高维 CSI 进行建模，忽视了信道在潜在空间中的低维结构；二是缺乏显式的空间一致性约束，使预测结果难以反映用户实际移动轨迹。因此，此类方法虽在短时预测中有效，但对通信系统中的空间决策支持能力有限[14]。

## 1.2 信道图谱及其在信道预测中的应用

信道图谱是一种面向无线信道的无监督表示学习方法，其目标是在未知用户位置信息的条件下，通过

CSI 样本之间的相似性关系构建低维嵌入空间，使该空间在一定程度上反映终端在物理空间中的几何结构。信道图谱的基本思想是：若两个用户处于相近的物理位置，其 CSI 具有较高相似性[5-6]。因此，通过定义合适的信道距离度量，可构建 CSI 样本之间的邻接关系，再通过降维方法生成信道图谱。常见的信道图谱实现包括基于多维尺度分析、t-分布随机近邻嵌入[8]、Isomap[15]及神经网络嵌入模型[16-17]等方法。

然而，现有信道图谱方法在建模过程中普遍忽略 CSI 的时间依赖性，仅通过几何邻接性构建嵌入空间，导致信道图谱在动态场景中稳定性不足。为提升预测性能，研究者通常在信道图谱输出结果上叠加插值或回归模型，但这类方法仍属于后处理策略，未在建模源头刻画信道的时序演化特性。此外，信道图谱插值方法依赖于局部线性假设和样本覆盖充分性，在用户快速移动或采样密度不均的情况下易出现误差积累，影响预测可靠性。因此，如何在维持信道图谱空间几何特性的同时，引入时间关联建模，仍是该领域亟待解决的关键问题。

### 1.3 深度时间序列模型在通信中的应用

近年来，随着深度学习技术的进步，研究者逐渐认识到，仅仅对时间序列进行点值预测已不足以有效刻画复杂系统的内在演化规律，更重要的是构建具有良好结构性的潜在表示空间，从而揭示数据背后的生成机制。因此，如何引入潜变量模型进行时序建模，已经成为当前研究的热点之一。AE 作为一种无监督学习方法，具有在数据中提取低维潜在特征的能力，能够有效地捕捉到数据的内在结构。传统的自编码器通过对数据进行编码与解码，学习出数据的低维表示。然而，AE 在处理时间序列数据时，缺乏对时间依赖性的建模，导致其难以在动态环境中保持较好的鲁棒性[18]。为了克服这一问题，近年来的研究将 RNN 引入到自编码器框架中，试图通过引入时序信息来改善潜空间的表达能力。LSTM 作为一种典型的 RNN 变体，因其独特的门控机制，能够有效地捕捉时间序列中的长期依赖关系，已广泛应用于语音识别、自然语言处理及时间序列预测等领域。

在此基础上，LSTM-AE 模型逐渐成为时序潜空间建模的重要方向[19-20]。该类模型通常将 LSTM 用作编码器或特征提取模块，将时间序列映射为低维潜空间的表示，并通过解码器重构原始信号，从而实现时间序列的建模与潜空间表示的联合优化。与传统自编码器或 LSTM 模型相比，LSTM-AE 能够同时捕捉时间序列中的长期依赖关系及潜空间的分布结构，从而提高了建模的稳定性与鲁棒性。

在无线通信领域，LSTM 及其变体已被成功应用于 CSI 序列的预测，并在多种场景中展现了比传统方法更优的性能[1, 4]。然而，这些方法通常仅仅对 CSI 进行回归建模，缺乏对信道样本之间几何结构的显式刻画，难以从空间层面对用户的分布进行建模。信道图谱作为一种无监督的几何表示学习方法，在无需位置信息的情况下能够构建低维信道图谱，具有较强的空间一致性刻画能力。然而，传统的信道图谱方法主要关注样本间的几何邻域保持，对时序一致性的建模相对薄弱，导致在动态环境下，潜空间的嵌入结果容易发生漂移。目前的研究尝试通过对信道图谱输出结果进行插值或简单的惯性建模来提升预测能力，但这些方法本质上仍是对已有嵌入结果的后处理，依赖于局部流形的平滑假设，难以有效刻画真实无线环境中的非线性变化及复杂的动态特性。在用户高速移动或环境发生频繁变化的场景下，这些插值方法容易出现误差积累，从而影响信道图谱的稳定性。

因此，当前的研究在“几何表示学习”和“时间序列建模”这两条技术路线之间依然存在割裂：一方面，信道图谱方法能够有效刻画空间结构，但缺乏对时序演化规律的建模；另一方面，LSTM 等模型具备时序建模优势，但忽略了信道在潜在空间中的几何特性。因此，如何在建模阶段统一空间表示与时间动态，是一个亟待深入探讨的问题。

针对这些不足，本文提出将 LSTM-AE 框架引入信道图谱过程中，既能够构建信道图谱，又能够刻画 CSI 的时间演化规律，实现动态信道流形的联合建模。与传统的“建图+插值”两阶段方法不同，所提出的方法在表征学习阶段直接引入时间一致性约束，使得潜空间同时满足几何邻近性与时间连续性，为构建稳定、可解释且具备预测能力的信道图谱提供了一种新的研究思路。

## 2 基于 LSTM-AE 的信道图谱构建方法

### 2.1 问题建模与总体框架

为建模信道状态信息中复杂时空依赖关系的捕捉过程，本文首先对原始 CSI 数据进行预处理和序列构建。原始 CSI 数据可表示为$\boldsymbol{g} \in \mathbb{R}^{N\times F}$，其中 $N$ 为数据点总数，$F$ 为 CSI 特征维度，包含天线阵列、子载波等信道参数。为充分利用 LSTM 网络的时序建模能力，并考虑到实际无线环境中用户位置的连续变化特性，将离散的 CSI 观测转换为连续时间序列，构建训练序列。对于每个时间点$i \geq L-1$，其中$L$为预设序列长度，

提取前$L$组连续时间步的 CSI 观测值构成输入序列$\boldsymbol{s}_i$=[$\boldsymbol{g}_{i-L+1}$,$\boldsymbol{g}_{i-L+2}$, ⋯ ,$\boldsymbol{g}_i$] ∈ $\mathbb{R}^{L\times N\times F}$。该序列对应的位置标签为末端时刻的真实位置$\boldsymbol{p}_i \in \mathbb{R}^{2\times \mathrm{N}}$，以此建立从历史信道状态到当前位置的映射关系。为平衡长期依赖捕捉与计算复杂度，经过实验验证，本文设定序列长度$L = 10$以在保持合理复杂度的同时确保时序信息的完整性。

经过上述预处理，元数据被转换为序列样本集合，其中每个样本包含一个固定长度（$L$）的连续时间段信道状态信息矩阵（由历史 CSI 向量行级拼接而成）及其对应的末端时刻空间位置标签，从而为后续 LSTM-AE 模型提供具备时空一致性的训练数据。

### 2.2 LSTM-AE 自编码器

为有效捕捉信道状态信息中的时序依赖关系，本文设计基于 LSTM 的时序编码器，如图 1 所示。LSTM 单元通过引入精密的“门控机制”来解决传统循环神经网络中的梯度消失和梯度爆炸问题，使其能够有效地学习长期依赖关系。LSTM 每个单元在每个时间步$t$的计算过程如下：

遗忘门控制历史信息的保留程度具体定义为：

矩阵、向量为粗斜体，除常量、数字、注释性单词外，其余改为斜体

$$f_t = \sigma\left(\boldsymbol{W}_f \cdot [\boldsymbol{h}_{t-1}, \boldsymbol{x}_t] + \boldsymbol{b}_f\right), \tag{1}$$

其中，**σ**为 sigmoid 激活函数，$W_f$和$b_f$分别为遗忘门的权重矩阵和偏置项，$h_{t-1}$为前一时刻的隐藏状态，$x_t$为当前时刻的输入。

输入门决定信息的更新程度：

$$i_t=\sigma(\boldsymbol{W}_i \cdot [\boldsymbol{h}_{t-1},\boldsymbol{x}_t]+\boldsymbol{b}_i), \tag{2}$$

其中，$W_{\boldsymbol{i}}$为输入门的权重矩阵，$b_{\boldsymbol{i}}$为输入门的偏置项。

候选记忆细胞生成当前时刻的潜在记忆定义为：

$$\tilde{\boldsymbol{c}}_t=\tanh(\boldsymbol{W}_c \cdot [\boldsymbol{h}_{t-1},\boldsymbol{x}_t]+\boldsymbol{b}_c), \tag{3}$$

其中，tanh为双曲正切激活函数，$W_{\boldsymbol{c}}$为候选记忆细胞的权重矩阵，$b_{\boldsymbol{c}}$为候选记忆细胞的偏置项。

记忆细胞更新结合遗忘与输入信息具体定义为：

$$\boldsymbol{c}_t=\boldsymbol{f}_t \odot \boldsymbol{c}_{t-1}+i_t \odot \tilde{\boldsymbol{c}}_t. \tag{4}$$

输出门控制当前隐藏状态的输出具体定义为：

$$o_t=\sigma(\boldsymbol{W}_\mathrm{o} \cdot [\boldsymbol{h}_{t-1},x_t]+\boldsymbol{b}_o), \tag{5}$$

$$\boldsymbol{h}_t=o_t \odot \tanh(c_t), \tag{6}$$

其中，$\boldsymbol{W_o}$为输出门的权重矩阵，$\boldsymbol{W_o}$为输出门的偏置项，$\boldsymbol{h}_t$为当前时刻的隐藏状态。

LSTM 编码器通过堆叠$U$个 LSTM 单元（本文中$U = 64$）逐步处理输入序列，最终获取序列级的特征表示为：

$$\boldsymbol{h}_\mathrm{enc}=\mathrm{LSTM}_\mathrm{encoder}(X) \in \mathbb{R}^U, \tag{7}$$

该隐藏状态 $\boldsymbol{h}_\mathrm{enc}$ 为整个输入序列的时序信息，为后续潜在空间构建提供丰富的特征基础。

为在保持信息完整性的同时实现维度约简，编码器输出经过全连接层映射至潜在空间：

$$\boldsymbol{z}=\boldsymbol{W}_z\boldsymbol{h}_\mathrm{enc}+\boldsymbol{b}_z \in \mathbb{R}^\mathrm{D}, \tag{8}$$

其中，$D = 32$为潜在空间维度，$W_{\boldsymbol{z}}$和$b_{\boldsymbol{z}}$分别为权重矩阵和偏置项。潜在空间$z$作为输入序列的紧凑表示，需同时满足两方面要求：一方面保留足够的信道状态信息以支持准确重建，另一方面具备良好的几何特性以支持位置嵌入。本文采用 ReLU 激活函数确保潜在表示的非线性变换能力：

$$\mathbf{z}_{\text{active}}=\text{ReLU}(\mathbf{z}). \tag{9}$$

基于潜在表示 $\mathbf{z}_{\text{active}}$，通过线性变换生成 2D 信道图谱坐标具体定义为：

$$\text{e}=\boldsymbol{W}_{\text{e}}\mathbf{z}_{\text{active}}+\boldsymbol{b}_{\text{e}}\in\mathbb{R}^2, \tag{10}$$

其中，$\boldsymbol{W}_{\text{e}}\in\mathbb{R}^{2\times D}$为嵌入权重矩阵，$\boldsymbol{b}_{\text{e}}\in\mathbb{R}^2$为偏置项。嵌入坐标 $e$ 直接对应于用户在物理空间中的位置估计，构成信道图谱的最终输出。

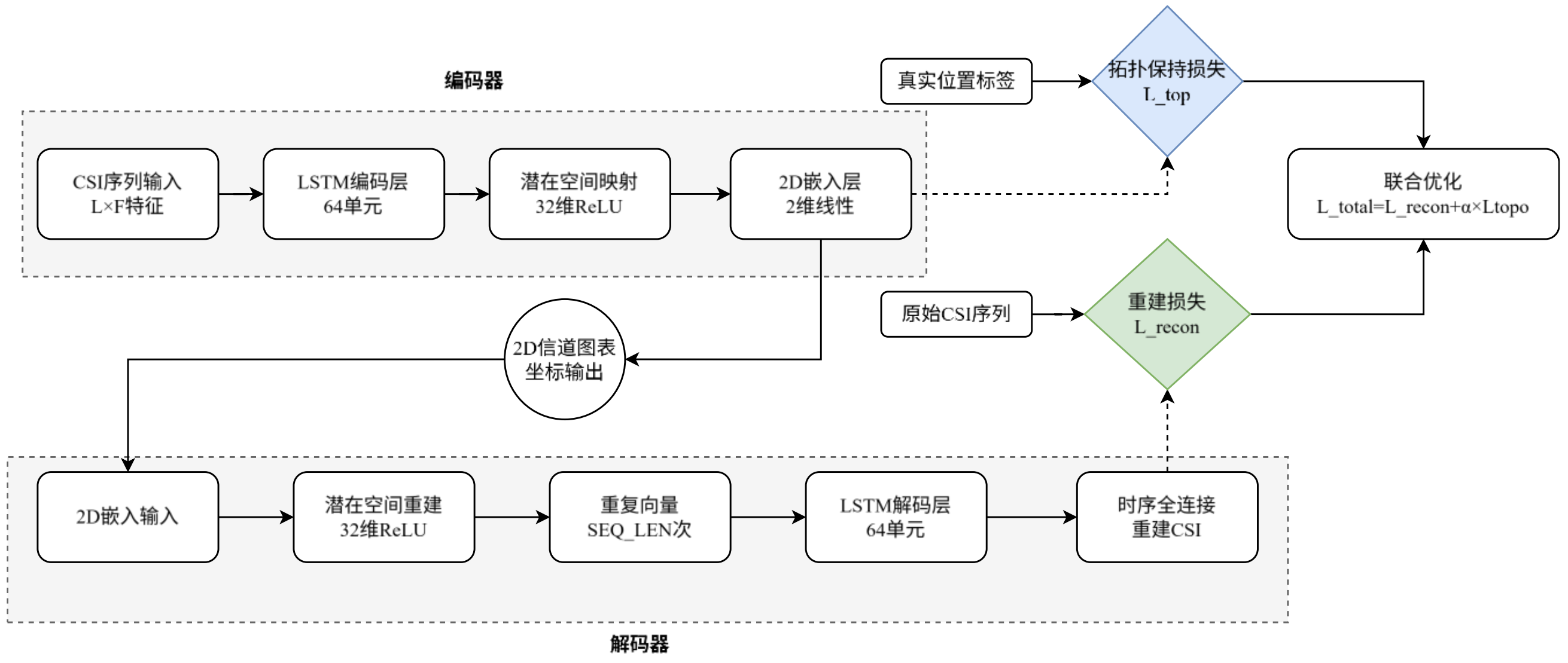

图 1　　LSTM-AE 模型

### 2.3　自编码器重建路径

为实现从低维潜在表示到高维 CSI 序列的精确重建，本文设计了与编码器对称的解码器结构。解码器首先通过全连接层将 2D 嵌入 $e$ 扩展至与编码器潜在空间相同的维度：

$$\mathbf{z}_{\text{dec}}=\boldsymbol{W}_{\text{dec}}\mathbf{z}_{\text{active}}+\boldsymbol{b}_{\text{dec}}\in\mathbb{R}^D, \tag{11}$$

其中，$\mathbf{z}_{\text{dec}}$表示解码器的初始潜在表示，$\boldsymbol{W}_{\text{dec}}\in\mathbb{R}^{D\times 2}$为解码器输入层的权重矩阵，$\boldsymbol{b}_{\text{dec}}\in\mathbb{R}^D$为偏置项，$D$=32 为潜在空间维度。为实现从单一潜在向量到时序序列的转换，采用 Repeat Vector 操作将潜在表示复制$L$次，构建解码器输入序列：

$$\mathbf{Z}_{\text{repeat}}=\left[\mathbf{z}_{\text{dec}},\mathbf{z}_{\text{dec}},\ldots,\mathbf{z}_{\text{dec}}\right]\in\mathbb{R}^{L\times D}. \tag{12}$$

LSTM 解码器以前述复制序列为输入，逐步生成重建序列：

$$\mathbf{H}_{\text{dec}}=\text{LSTM}_{\text{decoder}}\left(\mathbf{Z}_{\text{repeat}}\right)\in\mathbb{R}^{L\times U}, \tag{13}$$

其中，$\mathbf{H}_{\text{dec}}$表示解码器的隐藏状态序列，$U$为 LSTM 单元数量。最终，通过时间分布全连接层将解码器隐藏状态映射回原始 CSI 维度，表示完整的重建 CSI 序列的$\widehat{\boldsymbol{X}}$可表示为：

$$\widehat{\boldsymbol{X}}=[\widehat{\boldsymbol{x}}_1,\widehat{\boldsymbol{x}}_2,\ldots,\widehat{\boldsymbol{x}}_L]\in\mathbb{R}^{L\times F}, \tag{14}$$

其中，$\widehat{\boldsymbol{x}}_t=\boldsymbol{W}_{\text{out}}\mathbf{h}_{\text{dec},t}+\boldsymbol{b}_{\text{out}},t=1,2,\ldots,L$。

### 2.4　拓扑保持机制与损失函数

为精确重建 CSI 序列，并保证潜在空间具备良好的几何特性，能够反映真实的物理位置关系，本文引入基于成对距离的拓扑保持损失函数，其核心思想是最小化物理空间距离矩阵与嵌入空间距离矩阵之间的差异，表示拓扑保持损失的$\text{L}_{\text{topo}}$可以表示为：

$$L_{topo}=\frac{1}{N(N-1)}\sum_{i=1}^{N}\sum_{j\neq i}\left(\mathbf{d}_{ij}^{p}-\mathbf{d}_{ij}^{e}\right)^{2}, \tag{15}$$

其中，$N$为批次中的样本数量，$\mathbf{d}_{ij}^{p}$和$\mathbf{d}_{ij}^{e}$分别表示物理空间和嵌入空间中的成对距离。

为确保自编码器学习到有意义的潜在表示，我们采用均方误差（MSE）作为重建损失：

$$L_{recon}=\frac{1}{NLF}\sum_{n=1}^{N}\sum_{t=1}^{L}\sum_{f=1}^{F}\left(\boldsymbol{X}_{ntf}-\widehat{\boldsymbol{X}}_{ntf}\right)^{2}, \tag{16}$$

其中，$L_{recon}$ 表示重建损失，$N$ 为批次大小，$L$ 为序列长度，$F$ 为 CSI 特征维度，$X_{ntf}$表示第$n$个样本、第 t 个时间步、第 f 个特征的原始值，$\widehat{\boldsymbol{X}}_{ntf}$表示对应的重建值。

为实现拓扑保持与信号重建的双重目标，本文设计多目标联合损失$L_{total}$表示总损失函数为：

$$L_{total}=L_{recon}+\alpha L_{topo}, \tag{17}$$

其中$\alpha = 0.75$为拓扑保持损失的固定权重系数，通过实验验证该比例在本任务中表现最优。

3.5 信道图谱输出

基于训练完成的 LSTM-AE 模型，信道图谱的生成过程可以形式化表示为从 CSI 序列空间到 2D 嵌入空间的确定性映射$\mathcal{F}$可表示为：

$$\mathcal{F}: \mathbb{R}^{L\times F} \rightarrow \mathbb{R}^{2} \tag{18}$$

其中映射函数F由编码器网络参数化：

$$F_{\theta}(X)=f_{embed}\circ f_{latent}\circ f_{LSTM}(X), \tag{19}$$

其中，$\theta$表示编码器的所有参数，$f_{LSTM}$表示 LSTM 编码层，$f_{latent}$表示潜在空间映射层，$f_{embed}$表示 2D 嵌入层。

由于自编码器学习到的嵌入空间可能存在旋转、平移和缩放等几何变换，需要通过仿射变换将其与真实物理空间对齐。给定真实位置矩阵$\mathbf{P}=\left[\mathbf{p}_1,\mathbf{p}_2,\ldots,\mathbf{p}_N\right]^{T}\in\mathbb{R}^{N\times 2}$和嵌入矩阵$E$，我们寻求最优的仿射变换$\mathbf{T}^{*}$可表示为：

$$\boldsymbol{T}^{*}=\underset{\mathbf{T}}{\operatorname{argmin}}\ \|\ \boldsymbol{P}-\boldsymbol{E}T\ \|_{F}^{2}. \tag{20}$$

应用仿射变换得到对齐后的信道图谱$\boldsymbol{E}_{aligned}$可表示为：

$$\boldsymbol{E}_{aligned}=\boldsymbol{E}\boldsymbol{T}^{*}\in\mathbb{R}^{N\times 3}. \tag{21}$$

去除齐次坐标维度，得到最终的 2D 信道图谱$E_{final}$:

$$\boldsymbol{E}_{final}=\boldsymbol{E}_{aligned}[:,:2]\in\mathbb{R}^{N\times 2}, \tag{22}$$

由此最终得到图 2 所示的信道图谱潜空间图。

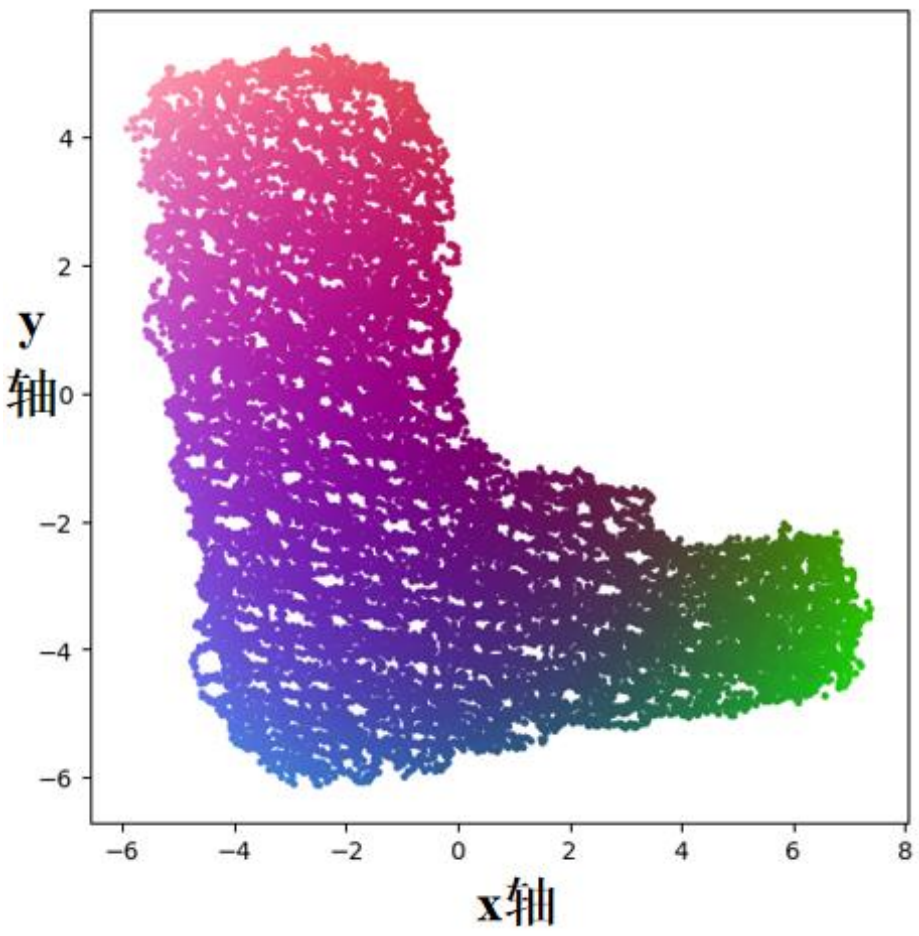

图 2　　LSTM-AE-信道图谱潜空间图

# 3　实验设置与结果分析

## 3.1　数据集描述

本文实验基于开源的工业室内环境下采集的分布式大规模 MIMO 信道数据，场景具有显著的多径散射与遮挡特性。用户在测量区域内以约 0.3 m/s 的低速连续移动，CSI 采样间隔为 0.192 s，属于低速、连续时序信道演化场景。所提出方法的性能验证主要适用于复杂室内或半封闭工业环境下的单用户低速移动场景，其在高速移动或纯室外宏蜂窝场景下的适用性仍有待进一步研究。具体而言，数据集包括了频域和时域的 CSI 样本，以及每个样本对应的用户位置标签。通过对原始数据进行处理与划分，生成了训练集和测试集，供模型进行训练和后续测试。以下是对数据集的详细描述。

（1）数据集选择

本文使用的数据集包含来自真实通信场景中不同用户的 CSI 数据和相应位置数据。为了测试所提出的基于 LSTM-AE 的信道图谱方法，数据集分训练集与验证集。训练集用于训练模型，而验证集则用于评估模型的性能，尤其是其在未知数据上的泛化能力。训练集包含时域和频域的 CSI 数据，以及对应的用户位置数据。训练数据集通过修正时间戳并映射有效索引，从原始数据中提取，确保数据的时序一致性。训练集的主要作用是让模型学习信道数据和用户位置之间的映射关系。测试集与训练集具有相似的结构，也包含时域和频域的 CSI 数据以及相应的用户位置数据。不同的是，测试集主要用于模型性能的测试，通过与真实位置的对比，评估模型在实际通信环境中的预测精度。选择该数据集的原因在于它能够有效地反映出无线通信系统中的时变信道特性，并且包含真实用户在不同移动轨迹下的 CSI 数据。这使得该数据集成为测试动态信道图谱建模方法的理想选择。

（2）数据集的特性

本实验所采用的信道状态信息（CSI）数据集，其核心参数配置如表 1 所示，该数据集针对无线信道建模的需求构建，具体特征与维度信息如下：

该数据集共包含 20827 个采样点，每个采样点对应 4 条天线链路（匹配 4×MIMO 的链路结构）；在 CSI 数据的构成上，每个采样点包含实部与虚部两类数据（对应复数通道维度为 2），且每个采样点配置 4 个子载波，每个子载波下设置 32 个 CSI 采样点，以刻画各信道的信号变化信息。数据集的原始维度为(20 827,4,2,4,13)，整合了样本数、天线链路、复数通道、子载波数及采样点数等多层维度信息；为适配后续定位任务与机器学习模型的输入要求，数据集经过载波频率偏移补偿与相位修正处理后，其修正后的形状调整为(20 827,4,2,4,32)，可直接用于后续模型的训练与测试流程。

表 1　数据集参数列表

| 特征 | 描述 | 维度 |
|---|---|---|
| 样本数 | 数据集中包含的采样点数量 | 20827 |
| 天线链路数 | 每个采样点对应的天线链路数（例如 4×MIMO 结构） | 4 |
| 复数通道 | 每个 CSI 采样点包含实部和虚部数据 | 2 |
| 子载波数量 | 每个 CSI 采样点使用的子载波数 | 4 |
| 子载波采样点数 | 每个子载波的 CSI 采样点数（每个信道的信号变化信息） | 32 |
| 数据集结构 | 数据集的整体结构，包含多个维度：样本数、天线链路、复数通道、子载波数和采样点数 | (20 827,4,2,4,13) |
| 数据集的修正后结构 | 数据集经过 CFO 补偿和相位修正后的形状，可以直接用于后续的定位任务或机器学习模型输入 | (20 827,4,2,4,32) |

（3）数据预处理

为了适应模型的输入要求，原始数据经过了以下几步预处理：

1）数据形状调整：初始数据的形状为(20 827,4,2,4,13)，其中包含多个维度。为了适配模型的输入需求，数据被展平成二维数据，维度为$(N,F)$，其中$N$为样本数，$F$为特征数。对于 CSI 数据，如果其维度大于 2，将被展平为二维数据。

2）滑动窗口切分：为了考虑时序信息，数据被切分为长度为 10 的序列（SEQ_LEN=10）。每个序列包含过去 10 个时间步的 CSI 数据，并且与当前时刻的真实位置数据配对。这样可以确保模型学习到 CSI 的时序依赖关系以及每个时刻对应的用户位置。

3）训练集与验证集划分：处理后的数据通过随机打乱后按 9:1 的比例划分为训练集和验证集。最终，训练集包含 18 736 个样本，验证集包含 2 082 个样本。其中数据集的输入包括多个维度，分别 CSI 数据和真实位置信息。

CSI 数据包含每个 CSI 样本包含时域或频域的信道状态信息，经过预处理后具有固定维度；真实位置为每个 CSI 样本对应的用户真实位置，作为训练时的目标标签。通过这种数据预处理与划分方式，模型能够学习到 CSI 数据与用户位置之间的关系，从而有效地进行信道图谱构建与位置预测任务。

### 3.2　实验设置

（1）模型配置

如表 2 所示，本实验所构建模型选用 Adam 作为优化器，初始学习率设定为 $e^{-3}$；模型结构层面，LSTM 单元数配置为 64，潜在空间维度（LATENT_DIM）与嵌入空间维度（EMBED_DIM）分别设定为 32 与 2。训练过程相关参数中，批量大小（BATCH_SIZE）设为 64，模型共执行 150 轮训练（EPOCHS）；损失函数权重方面，重建损失的权重固定为 1.0，拓扑损失的权重以超参数$\alpha$表示并依据实验需求调整。学习率调度采用 Reduce LR On Plateau 策略，对应的调度因子与耐心值分别设定为 0.5 与 5。数据集划分上，训练集样本数量为 18 736 个，验证集样本数量则为 2082 个。

表 2　模型配置列表

| 参数 | 设定 |
| --- | --- |
| 优化器 | Adam |
| 学习率 | $e^{-3}$ |
| LSTM 单元数（LSTM_UNITS） | 64 |
| 潜在空间维度（LATENT_DIM） | 32 |
| 嵌入空间维度（EMBED_DIM） | 2 |
| 批量大小（BATCH_SIZE） | 64 |
| 训练轮次（EPOCHS） | 150 |
| 重建损失权重（Reconstruction） | 1.0 |
| 拓扑损失权重（Embedding） | α |
| 学习率调度因子（Reduce LR On Plateau） | 0.5 |
| 学习率调度耐心值（Patience） | 5 |
| 训练集样本数（Train Samples） | 18736 |
| 验证集样本数（Validation Samples） | 2082 |

### 3.3　评估指标

在本实验中，为了全面评估所提出的基于 LSTM-AE 的信道图谱构建模型的性能，使用了四个关键评估指标：CT（Correlation of Topology）、TW（Topology Warping）、KS（Kullback-Leibler Divergence）以及 MAE（Mean Absolute Error）。以下对每个评估指标进行详细说明，并结合实验结果进行分析。

（1）模型效果评估：

如表 3 所示，对所提 LSTM-AE-信道图谱方法与基线信道图谱方法的性能进行量化评估，各指标的物理意义及对比结果分析如下：从拓扑结构匹配度来看，CT 指标衡量信道图谱与真实物理空间用户分布的相关性，TW 指标反映信道图谱的几何一致性，二者均以接近 1 为最优。所提 LSTM-AE-信道图谱方法的 CT 值达到 0.999 8、TW 值为 0.999 8，均趋近于理想值；相较于基线方法的 CT（0.996 6）与 TW（0.996 5），所提方法在拓扑相关性与几何一致性上实现了更精准的匹配，表明其生成的信道图谱能够更真实地复现物理空间的拓扑结构，且几何形变程度显著降低。在信道图谱的分布一致性方面，KS 指标用于度量模型输出与真实信道图谱的分布差异，以接近 0 为最优。所提方法的 KS 值仅为 0.0115，远小于基线方法的 0.083 5，说明该方法生成的信道图谱与真实数据的概率分布贴合度更高，有效缓解了传统方法中图表分布偏离真实信道特性的问题。从定位精度维度分析，MAE 衡量预测位置与真实位置的平均误差，值越小则定位性能越优。所提方法的 MAE 仅为 0.061 9 m，较基线方法的 0.484 7 m 实现了约 87%的误差降幅，充分测试了 LSTM-AE 时序建模机制对信道时变特性的捕捉能力，能够显著提升基于信道图谱的定位精度。

综合四类指标的对比结果可知，所提 LSTM-AE-信道图谱方法在拓扑匹配、几何一致性、分布贴合度及定位精度等多维度性能上均显著优于基线方法，测试了时序建模与低维表示学习融合框架的有效性。

表 3 模型效果评估指标（与基线对比）

| 方法 | CT↑ | TW↑ | KS↓ | MAE/m ↓ |
|---|---|---|---|---|
| 本文所提方法 | **0.999 8** | **0.999 8** | **0.011 5** | **0.061 9** |
| 信道图谱（基线） | 0.996 6 | 0.996 5 | 0.083 5 | 0.484 7 |

（2）误差向量对比分析

如图 3 所示，左图为基线方法在训练集上的误差可视化结果，图 3（b）为本文所提出方法的对应结果。可以观察到，基线模型在多数区域存在较为明显的预测偏移，真实位置与估计位置之间的错位较为普遍，特别是在空间边缘区域及轨迹弯折较大的位置处，误差向量较长，局部结构存在一定程度的畸变，表明其对信道动态变化的建模能力有限。

相比之下，本文方法所得到的信道图谱在整体结构上与真实空间分布高度一致，蓝色真实点与红色预测点高度重合，误差向量整体显著缩短，空间连续性和局部邻域关系保持良好，说明模型在嵌入空间中能够更准确地恢复真实几何拓扑结构。尤其是在轨迹变化复杂的高动态区域，所提方法仍能保持稳定的定位性能，未出现明显的结构扭曲或聚类塌陷现象，反映出较强的鲁棒性与泛化能力。

综上所述，本文提出的 LSTM-AE 模型在信道图谱构建与定位任务中表现出了优异的性能，能够有效捕捉信道的时变特性，并保持良好的空间一致性和定位精度。

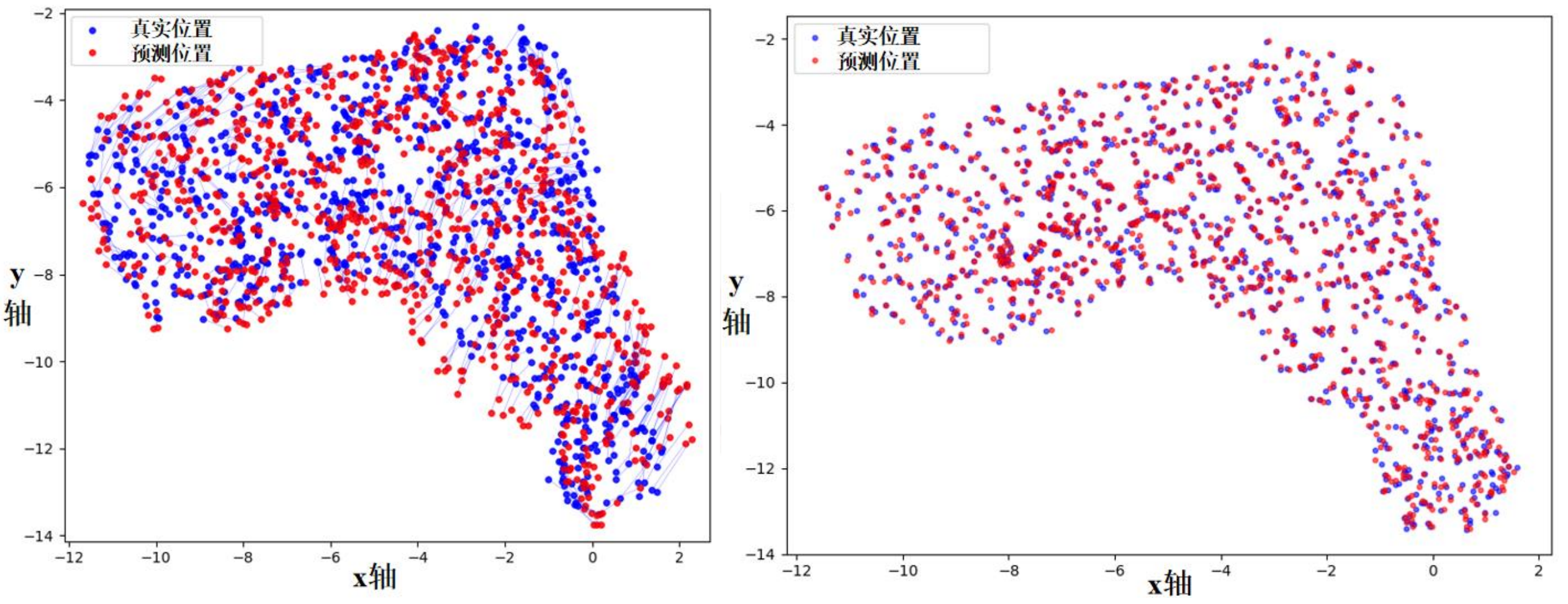

（a）几何不相似度驱动架构误差向量图 （b）本文提出的 LSTM-AE 架构误差向量图

图 3 不同方法构建的信道图谱在训练集上的误差向量对比

## 4 结束语

为解决复杂环境中信道图谱稳定性与拓扑一致性不足的问题，本文提出了一种基于 LSTM-AE 的时序信道图谱构建方法。信道时变特性与 CSI 过时问题是无线通信领域的核心挑战，现有研究如深度学习辅助的延迟容忍迫零预编码方案[21]、深度强化学习驱动的多用户波束成形设计[22]，均通过智能算法缓解信道老化对系统性能的影响，但在信道图谱的时序连续性建模与几何一致性协同优化方面仍有拓展空间。为此，本文在传统信道图谱框架基础上融入时序建模机制，通过 LSTM 网络捕捉 CSI 的时序依赖性，结合 AE 学习低维连续潜在表示，实现了信道几何一致性与时变特性建模的有机统一。实验结果表明，所提方法在真实通信场景中均表现出优异性能，特别是在信道图谱稳定性、轨迹连续性及长期预测能力方面，相较于传统信道图谱方法与现有部分智能优化方案[21-22]，具有显著优势，充分测试了时序建模与低维表示学习相结合的有效性。

本文仅聚焦于时序信道图谱的构建，未来将进一步拓展研究边界：一是探索时序信道图谱与当前移动通

信网络的结合，提升现有网络的性能，例如信道信息的获取[23]、资源分配[24]、MIMO 系统预编码设计[21]、边缘计算[25]等。二是信道图谱的利用依赖于用户位置信息的获取，结合高精度定位算法[26],[27]，信道图谱能够进一步提升移动通信网络的性能。

## 参考文献：